\documentclass[aps,pra,showpacs,twocolumn]{revtex4}

\usepackage{amsmath,epsfig,amsfonts,epsfig,graphicx
}

\begin{document}
\title{Fractional-Period Excitations in Continuum Periodic Systems}
\author{H. E. Nistazakis$^1$, Mason A. Porter$^2$, P. G. Kevrekidis$^3$, 
D. J. Frantzeskakis$^1$, A. Nicolin$^{4}$, and J. K. Chin$^5$}
\affiliation{$^1$ Department of Physics, University of Athens, Panepistimiopolis, Zografos, Athens 15784, Greece \\
$^2$ Department of Physics and Center for the Physics of Information, 
California Institute of Technology, Pasadena, CA  91125, USA \\
$^3$ Department of Mathematics and Statistics, University of Massachusetts, Amherst MA 01003-4515, USA \\ 
$^4$Niels Bohr Institute, DK-2100, Blegdamsvej 17, Copenhagen, Denmark \\
$^5$ Department of Physics, MFIT-Harvard Center for Ultracold Atoms, and Research Laboratory of 
Electronics, MIT, Cambridge, Massachusetts 02139, USA 
}

\begin{abstract}
We investigate the generation of fractional-period states in continuum 
periodic systems. As an 
example, we consider a 
Bose-Einstein condensate confined in an 
optical-lattice potential.  We show that when the potential is turned on 
non-adiabatically, the system explores a number of transient 
states whose periodicity is a fraction of that of the lattice. 
We illustrate the origin of fractional-period states analytically 
by treating them as 
resonant states of a parametrically forced Duffing oscillator 
and discuss their transient nature and potential observability. 
\end{abstract}

\maketitle

\section{Introduction}

In the past few years, there has been considerable interest 
in both genuinely discrete and continuum but periodic systems \cite{reviews}. 
These arise in diverse physical contexts \cite{focus}, including
coupled waveguide arrays and photorefractive crystals in nonlinear optics 
\cite{reviews1}, Bose-Einstein condensates (BECs) in
optical lattices (OLs) in atomic physics \cite{reviews2}, 
and DNA double-strand dynamics in biophysics \cite{reviews3}.  
One of the most interesting themes that emerges in this context
is the concept of ``effective discreteness'' induced by continuum periodic dynamics.  
There have been many efforts both to 
derive discrete 
systems that emulate the dynamics of continuum periodic ones 
\cite{derive} and to obtain continuum systems that mimic properties of discrete ones \cite{keenerus}. 
Additionally, the connection between discrete and continuum systems in various 
settings is one of the main research thrusts that has emerged from studies of the Fermi-Pasta-Ulam problem \cite{focus}.

This paper examines a type of excitation, not previously analyzed 
(to the best of our knowledge), with the intriguing characteristic that it can be observed
in continuum periodic systems but {\it cannot} be captured using a 
genuinely discrete description of the same problem. The reason for this 
is that these states bear the unusual feature that their length 
scale is a fraction of that of the continuum periodic potential. 
Thus, the ``fractional-period'' states reported in this paper are not 
stationary states of the latter problem, but rather transient excitations 
that persist for finite, observable times. 

To illustrate these fractional states, we consider the example of a trapped BEC in 
which an OL potential is turned on (as a non-adiabatic perturbation) \cite{reviews2}. 
Our results can also be applied in the context of optics by considering, 
for example, the effect of abruptly turning on an ordinary polarization beam
in a photorefractive crystal \cite{reviews1}. 
Our particular interest in BECs is motivated by recent 
experiments \cite{esslinger}, where after loading the condensate in an OL, 
the amplitude of the pertinent standing wave was modulated and the resulting 
excitations were probed. These findings were subsequently analyzed in the framework of the 
Gross-Pitaevskii equation in \cite{dalfovo1}, where it was argued that a parametric resonance 
occurs due to the OL amplitude modulation.  These results were 
further enforced by the 
analysis of \cite{dalfovo2}, which included a computation of the relevant 
stability diagram and growth rates of parametrically unstable modes.  The 
results of \cite{esslinger} were also examined in \cite{iucci} by treating 
the Bose gas as a Tomonaga-Luttinger liquid. 

A similar experiment, illustrating the controllability of such OLs, was recently reported in 
\cite{gemelke}, where instead of modulating the amplitude of the lattice, its location 
was translated (shaken) periodically.  This resulted in mixing between vibrational levels and 
the observation of period-doubled states. Such states were predicted earlier \cite{nicolin,map} 
in both lattice (discrete nonlinear Schr\"odinger) and continuum (Gross-Pitaevskii) 
frameworks in connection to a modulational instability \cite{smerzi,inguscio} and were also examined recently in \cite{seaman}. 
Period-multiplied states may exist as stationary (often unstable) solutions of such nonlinear problems and can usually be captured in the relevant lattice models.

To obtain fractional-period states, which cannot be constructed using Bloch's theorem \cite{ashcroft}, we will consider a setting similar to that of \cite{dalfovo1}, akin to the experiments of \cite{esslinger}. However, contrary to the aforementioned earlier
works (but still within the realm of the experimentally available
possibilities of, e.g., Ref.~\cite{esslinger}), we propose applying a strong 
non-adiabatic perturbation to the system (which originally consists of a magnetically confined BEC) 
by abruptly switching on an OL potential.  As a result, the BEC is far from its desired ground state.
Because of these nonequilibrium conditions, the system ``wanders'' in configuration space 
while trying to achieve its energetically desired state. In this process, we monitor the 
fractional-period states as observable transient excitations and report their signature in Fourier space. 
After presenting the relevant setup, we give an analysis of half-period and quarter-period states in a 
simplified setting.  We illustrate how
these states emerge, respectively, as harmonic and 1:2 superharmonic resonances 
of a parametrically forced Duffing oscillator describing the spatial dynamics of 
BEC standing waves (see Appendices A and B for details). 
We subsequently monitor these states in appropriately crafted numerical experiments and examine 
their dependence on system parameters. Finally, 
we also suggest possible means 
for observing the relevant states experimentally.

The rest of this paper is organized as follows. In Section II, we 
present the model and the analytical results.  (The details of the 
derivation of these results are presented in appendices; Appendix A 
discusses half-period states and Appendix B discusses quarter-period 
states.)  We present our numerical results in Section III and summarize 
our findings and present our conclusions in Section IV.

\section{Model and analysis}

\subsection{Setup}

A quasi-1D BEC is described by the dimensionless Gross-Pitaevskii (GP) equation \cite{reviews2,1d}, 
%
\begin{equation}
	i \frac{\partial \psi}{\partial t} = -\frac{1}{2} \frac{\partial^2\psi}{\partial x^2} 
+ g|\psi|^2 \psi + V(x,t)\psi\,, 
\label{ph1}
\end{equation}
where $\psi(x,t)$ is the 
mean-field wavefunction (with atomic density $|\psi|^2$ rescaled by the peak density $n_{0}$), 
$x$ is measured in units of the healing length $\xi=\hbar/\sqrt{n_{0} g_{1D}m}$ 
(where $m$ is the atomic mass), $t$ is measured in units of $\xi/c$ (where $c=\sqrt{n_{0}g_{1D}/m}$ 
is the Bogoliubov speed of sound), $g_{1D}=2\hbar \omega_{\perp} a$ is the effective 1D interaction strength, 
$\omega_{\perp}$ is the transverse confinement frequency, $a$ is the scattering length, 
and energy is measured in units of the chemical potential $\mu=g_{1D}n_{0}$. 
The nonlinearity strength $g$ (proportional to $a$)
is taken to be positive in connection to the $^{87}$Rb experiments of \cite{esslinger}. The potential,
\begin{equation}
	V(x,t)= \frac{1}{2}\Omega^2 x^2 + V_0 H(t) \left[ 1 + A \sin\left(\omega t \right) \right] \sin^2(q x)\,,
\label{ph2}
\end{equation}
%
consists of a harmonic (magnetic) trap of strength $\Omega \equiv \hbar \omega_{x}/g_{1D}n_{0}$ 
(where $\omega_{x}$ is the longitudinal confinement frequency) and an OL of 
wavenumber $q$, which is turned on abruptly at $t=0$ [via the Heaviside 
function $H(t)$]. The lattice depth, given by $V_{0} \left[ 1 + A \sin\left(\omega t \right) \right]$, is periodically modulated with frequency $\omega$.




Before the OL is turned on (i.e., for $ t< 0$), the magnetically 
trapped condensate is equilibrated in its ground state, which can be 
approximated reasonably well by 
the Thomas-Fermi (TF) cloud 
$u_{\mathrm{TF}}=\sqrt{\max \left\{ 0,\mu_{0}-V(x)\right\} }$, where 
$\mu_0$ is the normalized chemical potential \cite{reviews2}. 
The OL is then abruptly turned on and can be modulated weakly or strongly 
(by varying
$A$) and slowly or rapidly (by varying $\omega$).

To estimate the physical values of the parameters involved in this setting, we assume  
(for fixed values of the trap strength and normalized 
chemical potential, given by $\Omega=0.01$ and $\mu_{0}=1$, respectively) a magnetic trap 
with $\omega_{\perp}= 2\pi \times 1000\, {\rm Hz}$. Then, for a $^{87}$Rb ($^{23}$Na) condensate 
with 1D peak density $5 \times 10^7$ ${\rm m}^{-1}$ and longitudinal confinement frequency 
$\omega _{x}=2\pi \times 6 \,{\rm Hz}$ ($ 2 \pi \times 2.8\, {\rm Hz}$), the space and time 
units are $0.4\mu$m ($1.25\mu$m) and $0.27$ ms ($0.57$ ms), respectively, and the number 
of atoms (for $g=1$) is $N \approx 4200$ ($12000$).

\subsection{Analytical Results}

To provide an analytical description of fractional-period states, we initially consider the case of a homogeneous, untrapped condensate in a time-independent lattice (i.e., $\Omega=A=0$). 
We then apply a standing wave ansatz to Eq.~(\ref{ph1}) to obtain a parametrically forced 
Duffing oscillator (i.e., a cubic nonlinear Mathieu equation) describing the wavefunction's 
spatial dynamics. As examples, we analyze both half-period and quarter-period states. 
We discuss their construction briefly in the present section and provide further details 
in Appendices A and B, respectively.

We insert the standing wave ansatz
\begin{equation}
	\psi(x,t) = R(x)\exp\left(- i\mu_{0}t\right) \exp \left[ -i(V_{0}/2)t\right] 
\label{maw2}
\end{equation}
into Eq.~(\ref{ph1}) to obtain
\begin{equation}
	R'' + \delta R + \tilde{\alpha} R^3 + \varepsilon \tilde{V}_{0} R\cos(\kappa x) = 0\,,  
\label{ode}
\end{equation}
where 
primes denote differentiation with respect to $x$, $\delta = 2\mu_{0}$, 
$\tilde{\alpha} = -2g$, $\varepsilon \tilde{V}_{0} = V_{0}$, and $\kappa = 2q$.  

We construct fractional-period states using a multiple-scale perturbation expansion \cite{nayfeh}, 
defining $\eta \equiv \varepsilon x$ and $\xi \equiv bx = (1 + \varepsilon b_1 + \varepsilon^2 b_2 + \cdots)x$ for stretching parameters $b_j$. We then expand the wavefunction amplitude $R$ in a power series, 
\begin{equation}
	R = R_0 + \varepsilon R_1 + \varepsilon^2 R_2 + \varepsilon^3 R_3 + O(\varepsilon^4)\,. 
\label{eqr}
\end{equation}
Note that although $\xi$ and $\eta$ both depend on the variable $x$, the prefactor $\varepsilon$ in $\eta$ indicates that it varies much more slowly than $\xi$ so that the two variables describe phenomena on different spatial scales.  In proceeding with a perturbative analysis, we treat $\xi$ and $\eta$ as if they were independent variables (as discussed in detail in Ref.~\cite{nayfeh}) in order to isolate the dynamics arising at different scales \footnote{In many cases, one can make this procedure more mathematically rigorous (though less transparent physically) by examining the dynamics geometrically and introducing slow and fast manifolds.}.  We also incorporate a detuning into the procedure (in anticipation of our construction of resonant solutions) by also stretching the spatial dependence in the OL, which gives $W(\xi) = \varepsilon \tilde{V}_{0}\cos(\kappa \xi)$ for the last term in Eq.~(\ref{ode}) \cite{nlvibe}.  We insert Eq.~(\ref{eqr}) into Eq.~(\ref{ode}), expand the resulting ordinary differential equation (ODE) in a power series in $\varepsilon$, and equate the coefficents of like powers of $\varepsilon$.  

At each $O(\varepsilon^j)$, this yields a linear 
ODE in $\xi$ that $R_j = R_j(\xi,\eta)$ must satisfy:
%
\begin{equation}  
	L_\xi[R_j] \equiv  \frac{\partial^2 R_j}{\partial \xi^2} + \delta R_j = h_j(\xi,R_k,D^lR_k)\,, \label{oppy}
\end{equation}
where $h_j$ depends explicitly on $\xi$ and on $R_k$ and its derivatives (with respect to both $\xi$ and $\eta$) for all $k < j$.  We use the notation $D^lR_k$ in the right hand side of Eq.~(\ref{oppy}) to indicate its functional dependence on derivatives of $R_k$.  In particular, because of the second derivative term in Eq.~(\ref{ode}), these terms are of the form $\frac{\partial^2R_k}{\partial \xi^2}$, $\frac{\partial^2R_k}{\partial \eta^2}$, and $\frac{\partial^2R_k}{\partial \xi\partial\eta}$.  [See, for example, Eq.~(\ref{h1}) in Appendix A.]

We scale $\tilde{\alpha}$ (see the discussion below) to include at least one power of $\varepsilon$ in the nonlinearity coefficient in order to obtain an unforced harmonic oscillator when $j = 0$ (so that $h_0$ vanishes identically) \cite{note1}.  At each order, we expand $h_j$ in terms of its constituent harmonics, equate the coefficients of the independent secular terms to zero, and solve the resulting equations to obtain expressions for each of the $R_j$ in turn. (The forcing terms $h_j$ and the solutions $R_j$ are given in Appendix A for half-period states and Appendix B for quarter-period states.) 
Each $R_j$ depends on the variable $\eta$ through the integration constants obtained by integrating Eq.~(\ref{oppy}) with respect to $\xi$. The result of this analysis is an initial wavefunction, $\psi(x,0) = R(x)$, given by Eq.~(\ref{eqr}). 

We obtain half-period states of Eq.~(\ref{ode}) (and hence of the GP equation) by constructing 
solutions in harmonic (1:1) resonance with the OL (i.e., $\sqrt{\delta} = \kappa=2q$) \cite{mapsuper}. 
To perform the (second-order) multiple-scale analysis for this construction (see Appendix A), 
it is necessary to scale the nonlinearity to be of size $O(\varepsilon)$ 
(i.e., $\tilde{\alpha} = \varepsilon \alpha$), where $\varepsilon$ is a formal small parameter. 
[The OL is also of size $O(\varepsilon)$.]  We show below that full numerical simulations of 
the GP equation with a stationary OL using initial conditions obtained from the multiple-scale 
analysis yield stable half-period solutions even for large nonlinearities.  The oscillations in time about this state are just larger because of the $O(1)$ nonlinearity.  

We also obtain quarter-period states of Eq.~(\ref{ode}) (and hence of the GP equation) by constructing 
solutions in 1:2 superharmonic resonance with the OL (i.e., $\sqrt{\delta} = 2\kappa=4q$). 
Because the 1:2 superharmonic resonant solutions of the linearization of (\ref{ode}) 
[that is, of the linear Mathieu equation] are 4th-order Mathieu functions \cite{mathieu}, 
we must use a fourth-order multiple-scale expansion (see Appendix B) 
to obtain such solutions in the nonlinear problem when starting from trigonometric functions at $O(1)$. 
Accordingly, it is necessary to scale the nonlinearity to be of size $O(\varepsilon^4)$ 
(i.e., $\tilde{\alpha} = \varepsilon^4\alpha$).  [The OL is still of size $O(\varepsilon)$.]  
Nevertheless, as with half-period states, we show below that full numerical simulations of 
the GP equation with a stationary OL using initial conditions obtained from the multiple-scale 
analysis yield stable quarter-period solutions even for large nonlinearities.  The oscillations in time about this state are again larger because of the $O(1)$ nonlinearity.

\section{Numerical Results}

Having shown the origin of fractional-period states analytically, 
we now use numerical simulations to illustrate their dynamical relevance. 

\begin{figure}[tbp]
\vskip -0.28cm
\centering 
\includegraphics[width=3.75cm]{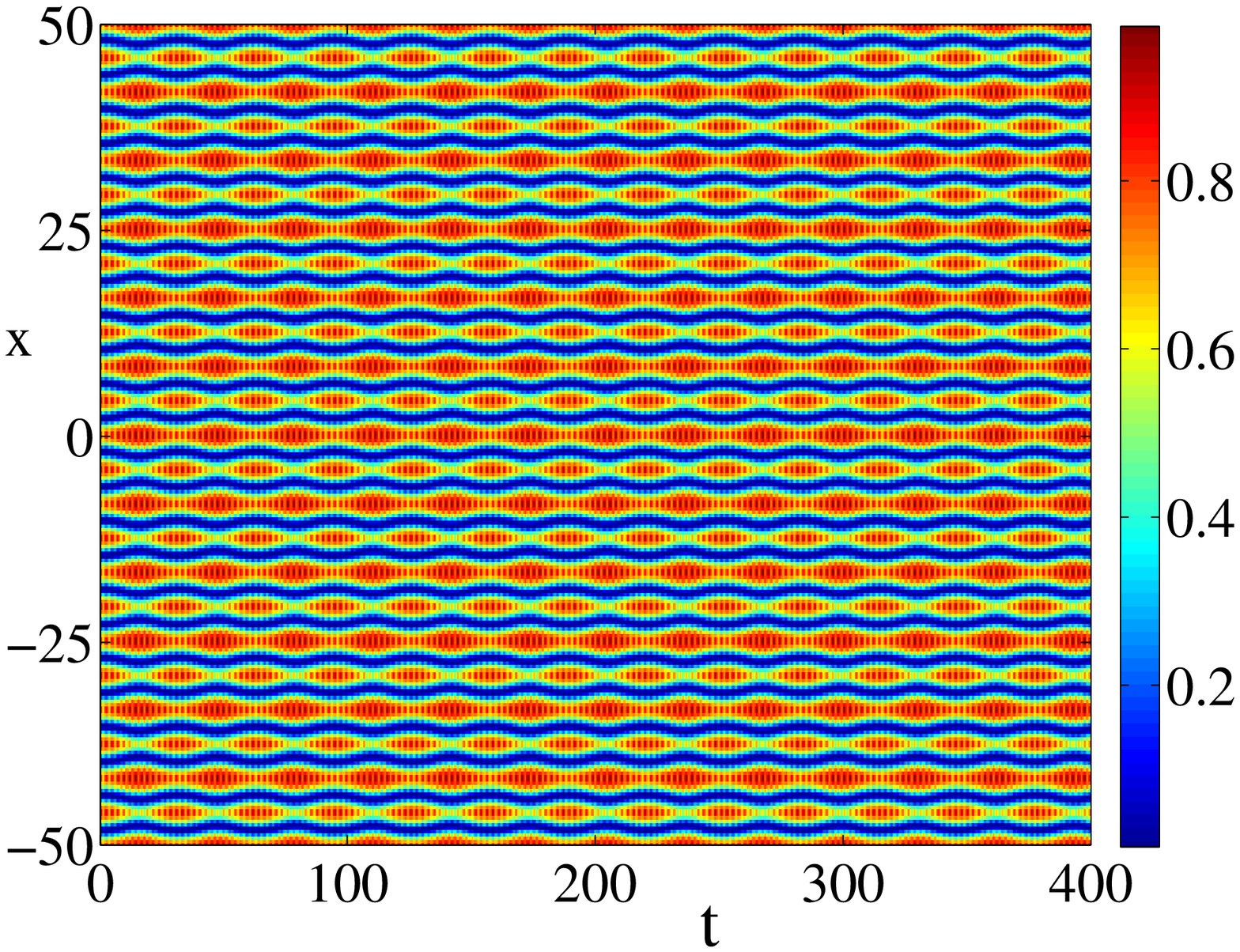}
\includegraphics[width=3.75cm]{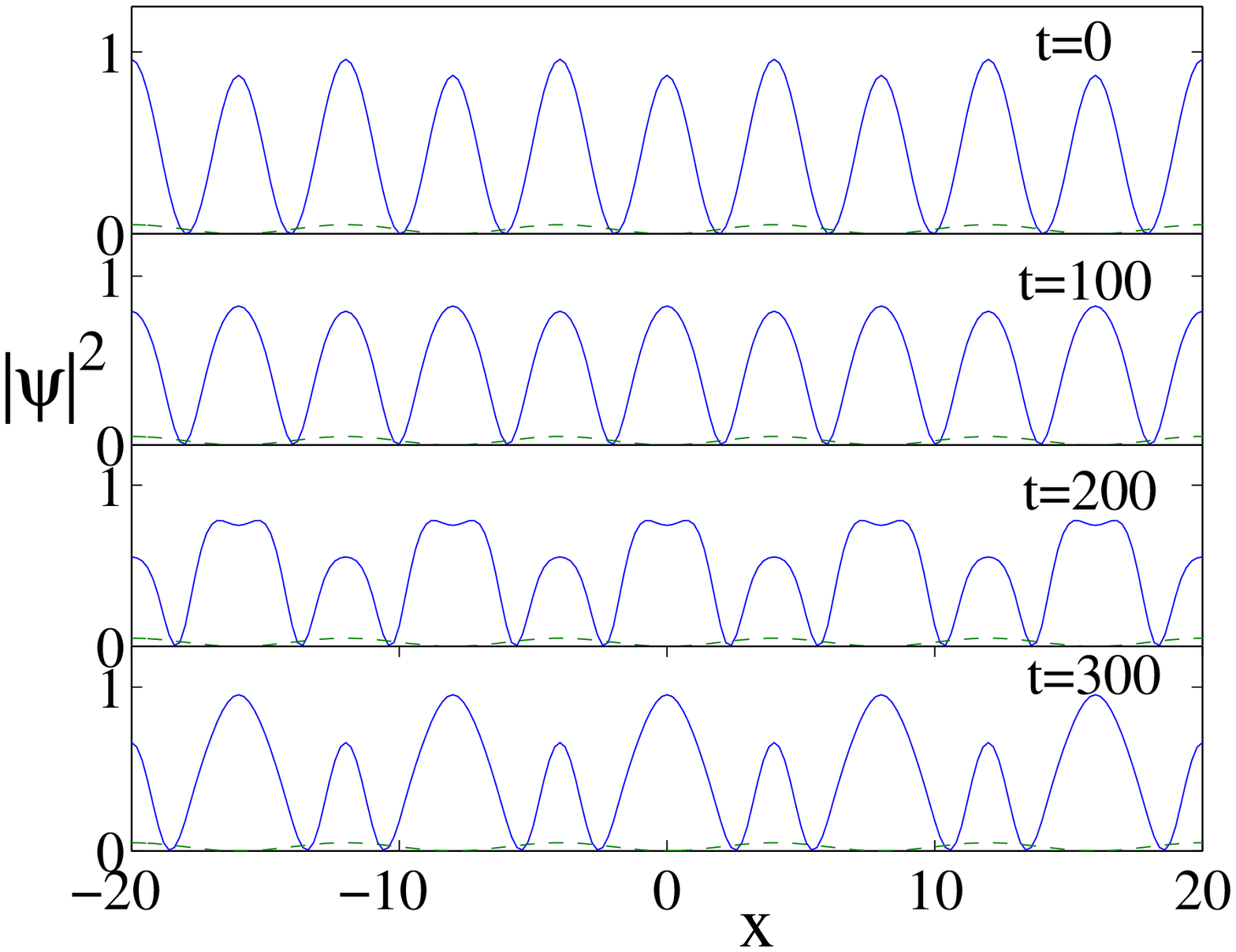}
\vskip -0.12cm
\caption{(Color online) Evolution of the half-period state found by numerically 
integrating Eq.~(\ref{ph1}). (Left) Space-time plot.  (Right) Snapshots of the density $|\psi |^2$ at $t=0$, $t = 100$, $t = 200$, and $t = 300$.  The parameter values are $\Omega=A=0$ (i.e., a time-independent OL and no magnetic trap), $g=1$, $\tilde{V}_{0}=1$, $q=\pi/8$ (so $\sqrt{\delta} = \pi/4$), $\varepsilon=0.05$, and $b_1 = b_2 = -1$. The dashed curve shows the OL potential.}
\label{Fig4}
\end{figure}

\begin{figure}[tbp]
\centering 
\includegraphics[width=7.5cm]{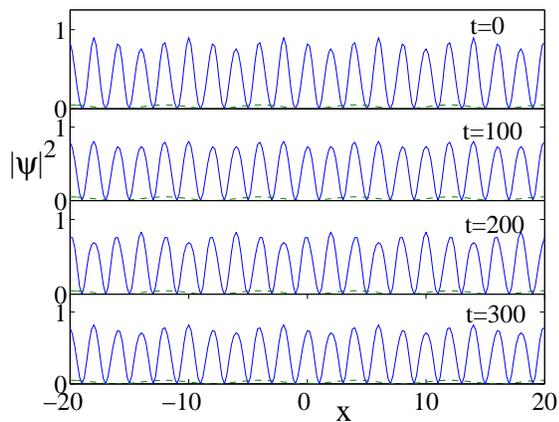}
\caption{(Color online) Evolution of the quarter-period state found by numerically integrating Eq.~(\ref{ph1}).  (As in Fig.~\ref{Fig4}, the lattice is time-independent and the magnetic trap is not included.)  We show snapshots of the density $|\psi |^2$ at $t=0$, $t=100$, $t=200$, and $t=300$.  The parameter values are $\Omega=A=0$, $g=1$, $\tilde{V}_{0}=1$, $q=\pi/8$ (so $\sqrt{\delta} = \pi/2$), $\varepsilon = 0.05$, and $b_1 = b_2 = b_3 = b_4 = -1$. The dashed curve shows the OL potential.}
\label{Fig4a}
\end{figure}

First, we consider the case with a stationary OL ($A = 0$) in the absence of the magnetic trap ($\Omega = 0$).  We examine the time-evolution of half-period states by numerically integrating 
Eq.~(\ref{ph1}) with the initial condition given by
Eqs.~(\ref{maw2}) and (\ref{eqr}).  We show an example in Fig.~\ref{Fig4}, where 
$g=1$, $\tilde{V}_{0}=1$, $q=\pi/8$, and $\varepsilon = 0.05$.
The half-period state persists for long times (beyond $t=300$). 
The parameter values correspond to a $^{87}$Rb ($^{23}$Na) BEC with a 1D peak density 
of $5 \times 10^7$ ${\rm m}^{-1}$ confined in a trap with frequencies 
$\omega_{\perp}= 2\pi \times 1000\, {\rm Hz}$, $\omega _{x}=0$, and number of atoms 
$N \approx 4000$ ($12500$). In real units, $t=300$ 
is about $80\, {\rm ms}$ ($170\, {\rm ms}$).  
We similarly examine the time-evolution of quarter-period states by numerically integrating 
Eq.~(\ref{ph1}) with the initial condition again given by Eqs.~(\ref{maw2}) and (\ref{eqr}), but now for the case of the superharmonic 1:2 resonance. 
In Fig.~\ref{Fig4a}, we show an example for a similar choice of parameters
as for half-period states \cite{note2}.

We subsequently examine the generation of such states through
direct numerical experiments using Eq.~(\ref{ph1}) in the presence of magnetic trapping.  Initially, we include the parabolic and periodic components of the potential but leave the potential 
time-independent, setting $A = 0$, $\mu_{0}=1$, $V_{0}=1$, and $q=\sqrt{2}/2$. 
First, we consider the case of a weak parabolic trap with $\Omega=0.001$. Using 
the nonlinearity coefficient $g=1$, we perform the numerical experiment as follows:  We 
integrate the GP equation in imaginary time to find the ``exact'' ground state 
(in the absence of the OL) and then we switch on the OL at $t = 0$. 
We 
monitor 
the density $|\psi(x,t)|^{2}$, its Fourier transform $|\Psi(k,t)|^{2}$, and the spectral components at 
$k=2q$ (the OL wavenumber) and $k=4q$ (half the wavenumber).
Note that the spectrum also contains a ``DC'' component (at $k=0$, corresponding to the ground state) 
as well as (very weak) higher harmonics.

\begin{figure}[tbp]
\centering \includegraphics[height=5cm,width=8cm]{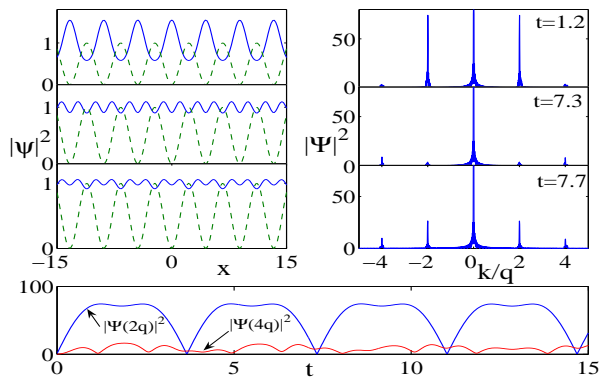}
\caption{
(Color online) Top panels: Snapshots of 
$|\psi(x,t)|^{2}$ (left) and its Fourier transform $|\Psi(k,t)|^{2}$ (right) for the case of a time-independent lattice and a magnetic trap at times $t=1.2$, $t=7.3$, and $t=7.7$.  The parameter values are $\Omega=0.001$, $\mu_{0}=1$, $V_{0}=1$, $q=\sqrt{2}/2$, and $A=0$.  The dashed curves (in the left panels) show the OL.  Bottom panel: Evolution of $|\Psi(4q)|^{2}$ (thin curve) and $|\Psi(2q)|^{2}$ (bold curve). For $t=1.2$, the density has the same period as the OL (i.e., $k=2q$).  For 
$t=7.3$ ($t=7.7$), we observe the formation of a quasi-harmonic (non-harmonic) half-period state with wavenumber $k=4q$.}
\label{Fig2}
\end{figure}

In general, the spectral component at $k=2q$ is significantly stronger than that at $k=4q$ 
(see the bottom panel of Fig.~\ref{Fig2}), implying that the 
preferable scale (period) of the system is 
set by the OL. This behavior is most prominent 
at certain times (e.g., at $t = 1.2$),
where the spectral component at $k=2q$ is much stronger than the other harmonics.
Nevertheless, there are specific time intervals 
(of length denoted by $\tau$) with $|\Psi(4q)|^{2} > |\Psi(2q)|^{2}$, where we observe the formation of 
what we will henceforth call a ``quasi-harmonic" half-period state.  For example, one can see such a state at $t = 7.3$. 
The purpose of the term ``quasi-harmonic" is to characterize half-period states whose second harmonic (at $k=4q$, in this case) is stronger than their first harmonic (at $k=2q$).  As mentioned above, such states have an almost sinusoidal shape, like the wavefunctions we constructed analytically. 

One can use the time-evolution of the spectral components 
as a quantitative method to identify the formation of half-period states. This 
diagnostic tool also reveals a ``revival'' of the state, 
which disappears and then reappears
a number of times before vanishing completely. Furthermore, we observe that 
other states that can also be characterized as half-period ones 
(which tend to have longer lifetimes than quasi-harmonic states) 
are also formed during the time-evolution, as shown in Fig. 3 at $t = 7.7$.   
These states, which we will hereafter call ``non-harmonic'' half-period states, have a shape which is definititvely non-sinusoidal (in contrast to the quasi-harmonic states); they are nevertheless periodic structures of period $k=4q$. 
In fact, the primary Fourier peak of the non-harmonic half-period states is always greater  
than the secondary one.  Such states can be observed for times $t$ such that 
the empirically selected condition of
$|\Psi(2q,t)|^2 \le  3 |\Psi(4q,t)|^2$ is satisfied. 

\begin{figure}[tbp]
\centering \includegraphics[width=7.5cm]{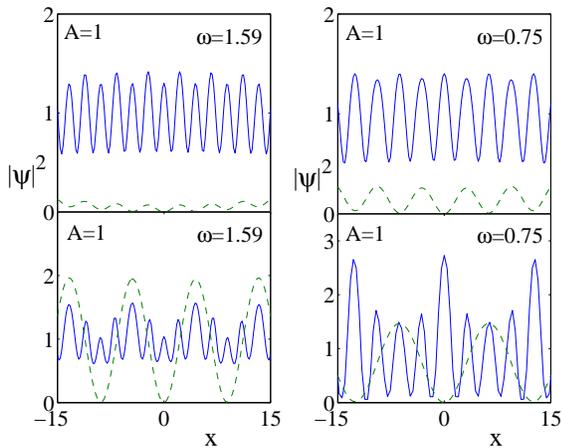}
\caption{
Half-period (top panels) and quarter-period (bottom panels) states for the case of a temporally modulated OL.  The state in the top left panel has wavenumber $q=\sqrt{2}/2$ (resonant), and that in the top right panel has wavenumber $q=1/2$ (non-resonant).  Similarly, the states in the bottom panels have wavenumbers $q=\sqrt{2}/4$ (left, resonant) and $q=1/4$ (right, non-resonant). 
} 
\label{Fig1}
\end{figure}

We next consider a stronger parabolic trap, setting $\Omega=0.01$. Because the system is generally 
less homogeneous in this case, we expect that the analytical prediction (valid for $\Omega=0$) 
may no longer be valid and that half-period states may 
cease to exist. We confirmed this numerically 
for the quasi-harmonic half-period states. However, non-harmonic half-period states do still appear.
The time-evolution of the spectral components at $k=2q$ and $k=4q$ is 
much more complicated and less efficient as a diagnostic tool, 
as $|\Psi(4q,t)|^{2} < |\Psi(2q,t)|^{2}$ for all $t$. Interestingly, 
the non-harmonic half-period states seem to persist as $\Omega$ is increased,
even when the resonance condition $\sqrt{\delta} = \kappa=2q$ is violated. 
For example, we found that for time-independent lattices (i.e., $A=0$), the lifetime of a half-period state  
in the resonant case with $q=\sqrt{2}/2$ (recall that $\mu_{0}=1$) 
is $\tau \approx 1.72$, whereas for $q=1/2$ 
it is $\tau \approx 0.84$. Moreover, the simulations show that 
the lifetimes become longer for periodically modulated OLs 
(using, e.g., $A=1$; also see the discussion below). In particular, in the 
aforementioned resonant (non-resonant) case with $q=\sqrt{2}/2$ ($q=1/2$), 
the lifetime of the half-period states has a maximum value,
at $\omega=1.59$ ($\omega=0.75$), of $\tau \approx 8.24$, or 
$4.7$ ms ($\tau \approx 5.72$, or $3.3$ ms) for a $^{23}$Na condensate. We show the formation of these 
states in the top panels of Fig.~\ref{Fig1}.

\begin{figure}[tbp]
\includegraphics[height=5cm,width=5cm]{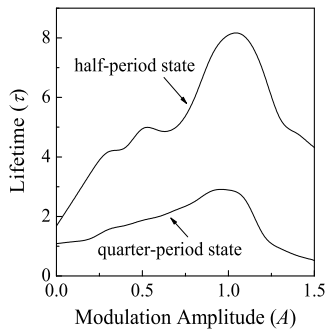}
\caption{The lifetime $\tau$ of the half-period (upper curve) and quarter-period (lower curve) states 
as a function of the lattice-modulation amplitude $A$. For the half-period 
(quarter-period) state, we show the case with wavenumber $q=\sqrt{2}/2$
($q=\sqrt{2}/4$).  In both examples, we used $\mu_{0}=1$ and lattice-modulation frequency $\omega=1.59$. 
} 
\label{lifetime}
\end{figure}

We also considered other fractional states. For example, using the same parameter 
values as before except for $\sqrt{\delta}=2 \kappa$ (so that $q=\sqrt{2}/4$), 
we observed quarter-period transient states 
with lifetime $\tau \approx 2.9$. These states occurred even in the 
non-resonant case with $q=1/4$ (yielding $\tau \approx 1$). We show these cases (for a temporally modulated lattice with modulation amplitude $A=1$) in 
the bottom panels of Fig.~\ref{Fig1}. In Fig.~\ref{lifetime}, we show the lifetime $\tau$ for 
the half-period and quarter-period states as a function of $A$. 
Observe that the lifetime becomes maximal (for values of $A \le 1.5$) around the 
value $A=1$ considered above.  Understanding the shape of these curves and the optimal 
lifetime dependence on $A$ in greater detail might be an interesting topic for further study.

\begin{figure}[tbp]
\includegraphics[height=3.5cm,width=6.5cm]{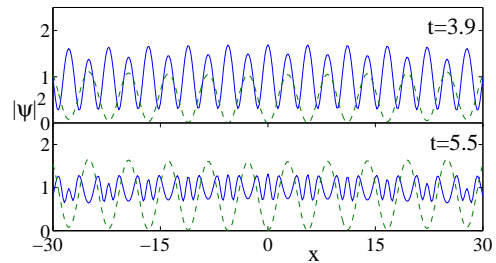}
\caption{(Color online) 
Density profiles for a fractional state with $\sqrt{\delta}=(5/4)\kappa$, so that $q=2\sqrt{2}/5$ 
(for $\mu_{0}=1$, lattice-modulation amplitude $A=1$, and lattice-modulation frequency $\omega=1.59$).  The system 
oscillates in time between half-period (top panel) and third-period (bottom panel) states.
} 
\label{Fig3}
\end{figure}

Finally, we also examined non-integer excitations, for which the system oscillates between the closest 
integer harmonics. For example, in Fig.~\ref{Fig3}, we show a state corresponding to $\sqrt{\delta}=(5/4)\kappa$, so that $q=2\sqrt{2}/5$ (with $\mu_{0}=1$). The system oscillates 
between the $k=4q$ (half-period) and $k=6q$ (third-period) states. 
Recall that the case presented in the top right panel of Fig.~\ref{Fig1} 
(with $q=1/2$) was identified as a ``non-resonant half-period state'' 
(the resonant state satisfies $q=\sqrt{2}/2$). 
Here it is worth remarking that this value, $q=1/2$, is closer to 
$q=\sqrt{2}/3$ (characterizing the third-period state)   
than to $\sqrt{2}/2$. Nevertheless, no matter which 
characterization one uses, the salient feature is that the value $q=1/2$ is 
nonresonant and lies between the third-period and half-period wavenumbers. 
Accordingly, the respective state oscillates between third-period and 
half-period states. Thus, in the case shown in the top right panel of 
Fig.~\ref{Fig1}, the third-period state also occurs (though we do 
not show it in the figure) and has a lifetime of $\tau=4.2$ ($2.4$ ms),
while for $q=2\sqrt{2}/5$ its lifetime is $\tau=2.92$ ($1.67$ ms). 
This indicates that states with wavenumbers closer to the value of the 
resonant third-period state have larger lifetimes.  This alternating oscillation between the nearest resonant period states is a typical feature that we have observed for the non-resonant cases.

\section{Conclusions}

In summary, we have investigated the formation and time-evolution of fractional-period 
states in continuum periodic systems. Although our analysis was based on a 
Gross-Pitaevskii equation describing Bose-Einstein condensates confined in optical lattices, 
it can also be applied to several other systems (such as photonic crystals in nonlinear optics). 
We have shown analytically and demonstrated numerically the formation of fractional-period
states and found that they may persist for sufficiently long times to be observed in experiments.
The most natural signature of the presence of such states should be available by monitoring the 
Fourier transform of the wavepacket through the existence of appropriate harmonics corresponding 
to the fractional-period states (e.g., $k=4q$ for half-period states, $k=8q$ for quarter-period states, etc.).

It would be interesting to expand the study of the parametric excitation of such states in order to better understand how to optimally select the relevant driving amplitude. Similarly, it would be valuable to examine more quantitatively the features of the ensuing states as a function of the
frequency of the parametric drive and the parabolic potential.


\vspace{2mm}

{\it Acknowledgements}: We thank Richard Rand for useful discussions and an anonymous referee for 
insightful suggestions. We also gratefully acknowledge support from the Gordon and Betty Moore 
Foundation (M.A.P.) and NSF-DMS-0204585, NSF-DMS-0505663 and 
NSF-CAREER (P.G.K.).

\section{Appendix A: Analytical Construction of Half-Period States}

To construct half-period states, we use the resonance relation 
$\sqrt{\delta} = \kappa$ and the scaling $\tilde{\alpha} = \varepsilon \alpha$, 
so that Eq.~(\ref{ode}) is written
\begin{equation}
	R'' + \kappa^2 R + \varepsilon\alpha R^3 + \varepsilon \tilde{V}_{0} R\cos(\kappa x) = 0  
\label{odeA}
\end{equation}
and Eq.~(\ref{oppy}) is written
\begin{equation}  
	L_\xi[R_j] \equiv  \frac{\partial^2 R_j}{\partial \xi^2} + \kappa^2 R_j = h_j(\xi,R_k,D^lR_k)\,, \label{oppyA}
\end{equation}
where we recall that $\eta = \varepsilon x$, $\xi = bx = (1 + \varepsilon b_1 + \varepsilon^2 b_2 + \cdots)x$, and $D^lR_k$ signifies the presence of derivatives of $R_k$ in the right hand side of the equation.  Because of the scaling in Eq.~(\ref{odeA}), $h_0 \equiv 0$, so that the $O(1)$ term is an 
unforced harmonic oscillator.  Its solution is 
\begin{equation}
	R_0 = A_0(\eta)\cos(\kappa\xi) + B_0(\eta)\sin(\kappa\xi)\,, \label{zero}
\end{equation}
where $A_0$ and $B_0$ will be determined by the solvability condition at $O(\varepsilon)$.

The $O(\varepsilon^j)$ ($j \geq 1$) equations arising from (\ref{odeA}) are forced harmonic 
oscillators, with  
forcing terms depending on the previously obtained $R_k(\xi,\eta)$ ($k < j$) and their derivatives. 
Their solutions take the form
\begin{equation}
	R_j = A_j(\eta)\cos(\kappa\xi) + B_j(\eta)\sin(\kappa\xi) + R_{jp}\,,
\end{equation}	 
where each $R_{jp}$ contains contributions from various harmonics.  As sinusoidal terms giving a 1:1 resonance with the OL arise at $O(\varepsilon^2)$, we can stop at that order.

At $O(\varepsilon)$, there is a contribution from both the OL and the nonlinearity, giving 
\begin{equation}
	h_1 = -\tilde{V}_0 R_0\cos(\kappa \xi) - \alpha R_0^3 - 2\frac{\partial^2R_0}{\partial\xi\partial\eta} - 2b_1\frac{\partial^2R_0}{\partial\xi^2}\,, \label{h1}
\end{equation}
where we recall that the OL depends on the stretched spatial variable $\xi $ because we are detuning from a resonant state \cite{nlvibe}.  With Eq.~(\ref{zero}), we obtain  
\begin{eqnarray}
	h_1 &=& \left[2b_1\kappa^2 A_0 - 2\kappa B_0' - \frac{3}{4}\alpha A_0(A_0^2 + B_0^2)\right]\cos(\kappa\xi) 
\nonumber \\
&+& \left[2b_1\kappa^2 B_0 + 2\kappa A_0' 
- \frac{3}{4}\alpha B_0(A_0^2 + B_0^2)\right]\sin(\kappa\xi) 
\nonumber \\
&+& \frac{\alpha A_0}{4}\left(-A_0^2 + 3B_0^2\right)\cos(3\kappa\xi)
\nonumber \\
&+& \frac{\alpha B_0}{4}\left(-3A_0^2 + B_0^2\right)\sin(3\kappa\xi) 
+ \frac{\tilde{V}_0A_0}{2} 
\nonumber \\
&+& \frac{\tilde{V}_0A_0}{2}\cos(2\kappa\xi) 
+ \frac{\tilde{V}_0B_0}{2}\sin(2\kappa\xi)\,. 
\label{o1}
\end{eqnarray}

For $R_1(\xi,\eta)$ to be bounded, the coefficients of the secular terms in Eq.~(\ref{o1}) 
must vanish \cite{nayfeh,nlvibe}.  The only secular harmonics are $\cos(\kappa\xi)$ and $\sin(\kappa\xi)$, and equating their coefficients to zero yields the following equations of motion describing the slow dynamics:
\begin{align}
        A_0' &= -b_1\kappa B_0 + \frac{3\alpha}{8 \kappa}B_0(A_0^2 + B_0^2) 
\,, \notag \\
        B_0' &= b_1\kappa A_0 - \frac{3\alpha}{8\kappa}A_0(A_0^2 + B_0^2) 
\,.   \label{slow1}
\end{align}
We convert (\ref{slow1}) to polar coordinates with $A_0(\eta) = C_0\cos[\varphi_0(\eta)]$ and $B_0(\eta) = C_0 \sin[\varphi_0(\eta)]$ and see immediately that each circle of constant $C_0$ is invariant.  The dynamics on each circle is given by 
\begin{equation}
  \varphi_0(\eta) = \varphi_0(0) + \left(b_1\kappa 
- \frac{3\alpha}{8\kappa}C_0^2\right)\eta\,.
\end{equation}
We examine the special circle of equilibria, corresponding to periodic orbits of (\ref{odeA}), which satisfies
\begin{equation}
        C_0^2 = A_0^2 + B_0^2 = \frac{8b_1\kappa^2}{3\alpha}\,. \label{echo}
\end{equation}
In choosing an initial configuration for numerical simulations of the GP equation (\ref{ph1}), we set $B_0=0$ without loss of generality.

Equating coefficients of (\ref{oppyA}) at $O(\varepsilon^2)$ yields
\begin{equation}
        \frac{\partial^2 R_2}{\partial \xi^2} + \kappa^2 R_2 = h_2\,, 
\end{equation}
where the forcing term again contains contributions from both the OL and the nonlinearity:
\begin{align}        
        h_2 &= -(b_1^2 + 2b_2)\frac{\partial^2R_0}{\partial \xi^2} - \frac{\partial^2R_0}{\partial \eta^2} - 2b_1\frac{\partial^2 R_0}{\partial \xi \partial \eta} \notag \\ &\quad - 3\alpha R_0^2R_1 - 2b_1\frac{\partial^2 R_1}{\partial \xi^2} - 2\frac{\partial^2 R_1}{\partial \xi \partial \eta}  - R_1\tilde{V}_0\cos(\kappa\xi) \,. \label{2eq}
\end{align}
Here, one inserts the expressions for $R_0$, $R_1$, and their derivatives into the function $h_2$.

To find the secular terms in Eq.~(\ref{2eq}), we compute
\begin{align}
        R_1(\xi,\eta) &= A_1(\eta)\cos(\kappa\xi) + B_1(\eta)\sin(\kappa\xi) + R_{1p}(\xi,\eta)\,, \notag \\
        R_{1p}(\xi,\eta) &= c_1\cos(3\kappa\xi) + c_2\sin(3\kappa\xi) \notag \\ &\quad + c_3 + c_4\cos(2\kappa\xi) + c_5\sin(2\kappa\xi)  \,, \label{r1}
\end{align}
where
\begin{align}
        c_1 &= \frac{\alpha}{32\kappa^2}A_0(A_0^2 - 3B_0^2)\,, \quad
        c_2 = \frac{\alpha}{32\kappa^2}B_0(3A_0^2 - B_0^2)\,, \notag \\
        c_3 &= -\frac{\tilde{V}_0A_0}{2\kappa^2}\,, \quad
        c_4 = \frac{\tilde{V}_0A_0}{6\kappa^2}\,, \quad
        c_5 = \frac{\tilde{V}_0B_0}{6\kappa^2}\,.
\end{align}

After it is expanded, the function $h_2$ contains harmonics of the form $\cos(0 \xi) = 1$, 
$\cos(\kappa \xi)$ (the secular terms), $\cos(2\kappa \xi)$, $\cos(3\kappa \xi)$, 
$\cos(4\kappa \xi)$, and $\cos(5\kappa \xi)$ (as well as sine functions with the same arguments). Equating the secular cofficients to zeros gives the following equations describing the slow dynamics:
\begin{align}
        A_1' &= \frac{1}{3072\kappa^5}\left[\left( f_1(\alpha,\kappa)B_0^2 + f_2(\alpha,\kappa)A_0^2 + f_3(\alpha,\kappa,b_1) \right)B_1 \right. \notag \\ &\quad \left.+ f_4(\alpha,\kappa)A_0B_0A_1 + f_5(\alpha,\kappa)B_0^5 \right. \notag \\ &\quad \left. + f_6(\alpha,\kappa)A_0^2B_0^3 + f_7(\alpha,\kappa)A_0^4B_0 + f_{8s}(\alpha,\kappa,b_2)B_0 \right] \,, \notag \\
        B_1' &= \frac{1}{3072\kappa^5}\left[\left( f_1(\alpha,\kappa)A_0^2 + f_2(\alpha,\kappa)B_0^2 + f_3(\alpha,\kappa,b_1) \right)A_1 \right. \notag \\ &\quad \left. + f_4(\alpha,\kappa)A_0B_0B_1 + f_5(\alpha,\kappa)A_0^5 \right. \notag \\ &\quad \left.  + f_6(\alpha,\kappa)A_0^3B_0^2 + f_7(\alpha,\kappa)A_0B_0^4 + f_{8c}(\alpha,\kappa)A_0 \right] \,, \label{resharm}
\end{align}
where
\begin{align}
        f_1(\alpha,\kappa) &= 3 f_2(\alpha,\kappa)   \,, \notag \\
        f_2(\alpha,\kappa) &= -1152\alpha\kappa^4 \,, \notag \\
        f_3(\alpha,\kappa,b_1) &= 3072\kappa^6b_1 \,, \notag \\
        f_4(\alpha,\kappa) &= 2 f_2(\alpha,\kappa)  \,, \notag \\
        f_5(\alpha,\kappa) &= 180\alpha^2\kappa^2 \,, \notag \\
        f_6(\alpha,\kappa) &= 2 f_5(\alpha,\kappa)  \,, \notag \\
        f_7(\alpha,\kappa) &= f_5(\alpha,\kappa)  \,, \notag \\
        f_{8s}(\alpha,\kappa,b_2) &= f_{non}(\alpha,\kappa) - 128\tilde{V}_0^2\kappa^2\,, \notag \\ 
        f_{8c}(\alpha,\kappa) &=  f_{non}(\alpha,\kappa) + 640\tilde{V}_0^2\kappa^2\,, \notag \\ 
        f_{non}(\alpha,\kappa) &= 3072\kappa^6b_2 \,. \label{harmf}
\end{align}
We use the notation $f_{non}$ to indicate the portions of the quantities $f_{8s}$ and $f_{8c}$ that arise from non-resonant terms.  The other terms in these quantities, which depend on the lattice amplitude $V_0$, arise from resonant terms.

Equilibrium solutions  of (\ref{resharm}) satisfy 
\begin{widetext}
\begin{align}
        A_1 &= \frac{(f_1B_0^2 + f_2A_0^2 + f_3)(f_5A_0^5 + f_6A_0^3B_0^2 + f_7A_0B_0^4 + f_{8c}A_0) - (f_4A_0B_0)(f_5B_0^5 + f_6A_0^2B_0^3 + f_7A_0^4B_0 + f_{8s}B_0)}{f_4^2A_0^2B_0^2 -(f_1B_0^2 + f_2A_0^2 + f_3)(f_1A_0^2 + f_2B_0^2 + f_3)} \,, \notag \\
        B_1 &= \frac{(f_1A_0^2 + f_2B_0^2 + f_3)(f_5B_0^5 + f_6A_0^2B_0^3 + f_7A_0^4B_0 + f_{8s}B_0) - (f_4A_0B_0)(f_5A_0^5 + f_6A_0^3B_0^2 + f_7A_0B_0^4 + f_{8c}A_0)}{f_4^2A_0^2B_0^2 -(f_1B_0^2 + f_2A_0^2 + f_3)(f_1A_0^2 + f_2B_0^2 + f_3)}\,, \label{order}
\end{align}
\end{widetext}
where one 
uses an equilibrium value of $A_0$ and $B_0$ from Eq.~(\ref{echo}). 
Inserting equilibrium values of $A_0$, $B_0$, $A_1$, and $B_1$ into Eqs.~(\ref{zero}) and (\ref{r1}), we obtain the spatial profile $R = R_0 + \varepsilon R_1 + O(\varepsilon^2)$ used as the initial wavefunction in the numerical simulations of the full GP equation (\ref{ph1}) with a stationary OL.

\section{Appendix B: Analytical Construction of Quarter-Period States}

To construct quarter-period states, we use the resonance relation $\sqrt{\delta} = 2\kappa$ and 
the scaling $\tilde{\alpha} = \varepsilon^4 \alpha$, so that Eq.~(\ref{ode}) is written
\begin{equation}
	R'' + 4\kappa^2 R + \varepsilon^4\alpha R^3 + \varepsilon \tilde{V}_{0} R\cos(\kappa x) = 0  
\label{odeB}
\end{equation}
and Eq.~(\ref{oppy}) is written
\begin{equation}  
	L_\xi[R_j] \equiv  \frac{\partial^2 R_j}{\partial \xi^2} + 4\kappa^2 R_j = h_j(\xi,R_k,D^lR_k)\,, \label{oppyB}
\end{equation}
where $\eta = \varepsilon x$ and $\xi = bx = (1 + \varepsilon b_1 + \varepsilon^2 b_2 + \cdots)x$, as before.

Because of the scaling in (\ref{odeB}), $h_0 \equiv 0$ (as in the case of half-period states), 
so that the $O(1)$ term is an unforced harmonic oscillator.  It has the solution 
\begin{equation}
	R_0 = A_0(\eta)\cos(2\kappa\xi) + B_0(\eta)\sin(2\kappa\xi)\,, \label{zeroB}
\end{equation}
where $A_0^2 + B_0^2 = C_0^2$ is an arbitrary constant (in the numerical simulations, 
we take $B_0 = 0$ without loss of generality). With the different scaling of the 
nonlinearity coefficient, the value $C_0^2$ is not constrained as 
it was in the case of half-period states
(see Appendix A).

The $O(\varepsilon^j)$ ($j \geq 1$) equations arising from (\ref{oppyB}) are forced harmonic 
oscillators, with forcing terms depending on the previously obtained $R_k(\xi,\eta)$ 
($k < j$) and their derivatives. Their solutions take the form
\begin{equation}
	R_j = A_j(\eta)\cos(2\kappa\xi) + B_j(\eta)\sin(2\kappa\xi) + R_{jp}\,,
\end{equation}	 
where $R_{jp}$ contain contributions from various harmonics. As sinusoidal terms 
giving a 1:2 resonance with the OL arise at $O(\varepsilon^4)$, we can stop at that order.

The equation at $O(\varepsilon)$ has a solution of the form
\begin{equation}
  	R_1 = A_1(\eta)\cos(2\kappa\xi) + B_1(\eta)\sin(2\kappa\xi) + R_{1p}\,.
\end{equation}
The coefficients $A_1$ and $B_1$ are determined using a solvability condition obtained at $O(\varepsilon^2)$ by requiring that the secular terms of $h_2$ vanish.  This yields
\begin{align}
  	A_1 &= \frac{\tilde{V}_0}{16b_1\kappa^2}(c_{11} + c_{12}) - \frac{b_2}{b_1}A_0\,, \notag \\
  	B_1 &= \frac{\tilde{V}_0}{16b_1\kappa^2}(c_{13} + c_{14}) - \frac{b_2}{b_1}B_0\,.
\end{align}
The particular solution is
\begin{equation}
  	R_{1p} = c_{11}\cos(\kappa\xi) + c_{12}\cos(3\kappa\xi) 
+ c_{13}\sin(3\kappa\xi) + c_{14}\sin(\kappa\xi)\,, \label{1B}
\end{equation}
where
\begin{align}
  	c_{11} &= -\frac{\tilde{V}_0A_0}{6\kappa^2} \,, \quad
  	c_{12} = \frac{\tilde{V}_0A_0}{10\kappa^2} \,, \notag \\
  	c_{13} &= \frac{\tilde{V}_0B_0}{10\kappa^2} \,, \quad
  	c_{14} = -\frac{\tilde{V}_0B_0}{6\kappa^2} \,.
\end{align}

The solution at $O(\varepsilon^2)$ has the form
\begin{equation}
  	R_2 = A_2(\eta)\cos(2\kappa\xi) + B_2(\eta)\sin(2\kappa\xi) + R_{2p}(\xi,\eta)\,. \label{2B}
\end{equation}
The coefficients $A_2$ and $B_2$ are determined using a solvability condition obtained at $O(\varepsilon^3)$ by requiring that the secular terms of $h_3$ vanish.  This yields
\begin{align}
  	A_2 &= \frac{\tilde{V}_0}{16b_1\kappa^2}(c_{21} + c_{22}) - \frac{\tilde{V}_0}{32\kappa^2}(c_{11} + c_{12}) \notag \\ &\quad - \frac{b_3}{b_1}A_0 - \frac{b_2}{b_1}A_1 - \frac{\tilde{V}_0^2A_0}{480\kappa^4} \,, \notag \\
  B_2 &= -\frac{\tilde{V}_0}{16b_1\kappa^2}(c_{23} + c_{24}) - \frac{\tilde{V}_0}{32\kappa^2}(c_{13} + c_{14}) \notag \\ &\quad - \frac{b_3}{b_1}B_0 - \frac{b_2}{b_1}B_1 - \frac{\tilde{V}_0^2B_0}{480\kappa^4} \,.
\end{align}
The particular solution is
\begin{eqnarray}
  	R_{2p} &=& c_{21}\cos(\kappa\xi) + c_{22}\cos(3\kappa\xi) + c_{23}\sin(3\kappa\xi) \nonumber \\ 
&+& c_{24}\sin(\kappa\xi) 
+ c_{25}\cos(4\kappa\xi) \nonumber \\ 
&+& c_{26} + c_{27}\sin(4\kappa\xi)\,, \label{r2p}
\end{eqnarray}
where
\begin{align}
  	c_{21} &= -\frac{\tilde{V}_0A_1}{6\kappa^2} + \frac{b_1\tilde{V}_0A_0}{9\kappa^2}\,, \notag 
\\
  	c_{22} &= \frac{\tilde{V}_0A_1}{10\kappa^2} - \frac{3b_1\tilde{V}_0A_0}{25\kappa^2} \,, 
\notag \\
  	c_{23} &= \frac{\tilde{V}_0B_1}{10\kappa^2} - \frac{3b_1\tilde{V}_0B_0}{25\kappa^2} \,, 
\notag \\
  	c_{24} &= -\frac{\tilde{V}_0B_1}{6\kappa^2} + \frac{b_1\tilde{V}_0B_0}{9\kappa^2} \,, \notag 
\\
  	c_{25} &= \frac{\tilde{V}_0^2A_0}{240\kappa^4} \,, \quad
  	c_{26} = \frac{\tilde{V}_0^2A_0}{48\kappa^4} \,, \quad
  	c_{27} = \frac{\tilde{V}_0^2B_0}{240\kappa^4} \,. 
\end{align}
Note that the harmonics $\cos(0\xi)$ and $\sin(0\xi)$ occur in (\ref{r2p}) and are reduced appropriately.  
(The arguments of this sine and cosine arise because of our particular resonance relation.)  


At $O(\varepsilon^3)$, we obtain solutions of the form
\begin{equation}
  	R_3 = A_3(\eta)\cos(2\kappa\xi) + B_3(\eta)\sin(2\kappa\xi) + R_{3p}(\xi,\eta)\,. \label{3B}
\end{equation}
The coefficients $A_3$ and $B_3$ are determined using a solvability condition obtained at $O(\varepsilon^4)$ by requiring that the secular terms of $h_4$ vanish.  Because of the scaling in (\ref{odeB}), the effects of the nonlinearity manifest in this solvability condition.  The resulting coefficients are
\begin{eqnarray}
A_3 =&-&\frac{\tilde{V}_0^4A_0}{3375b_1\kappa^8} - \frac{b_3}{b_1}A_1 - \frac{\tilde{V}_0^2b_2A_0}{1800b_1\kappa^4} 
\nonumber \\ 
&-& \frac{\tilde{V}_0^2A_1}{1800\kappa^4} - \frac{19b_1\tilde{V}_0^2A_0}{54000\kappa^4} 
+ \frac{3\alpha A_0^3}{32b_1\kappa^2} 
\nonumber \\ 
&-& \frac{b_2}{b_1}A_2 - \frac{b_4}{b_1}A_0 - \frac{\tilde{V}_0^2A_2}{240b_1\kappa^4} + \frac{3\alpha A_0B_0^2}{32b_1\kappa^2} 
\label{bb1}
\end{eqnarray}

\begin{eqnarray}  	
B_3 = &-&\frac{b_2}{b_1}B_2 + \frac{3\alpha A_0^2B_0}{32b_1\kappa^2} - \frac{b_4}{b_1}B_0 
- \frac{\tilde{V}_0^2B_2}{240b_1\kappa^4} 
\nonumber \\
&-& \frac{\tilde{V}_0^2B_1}{1800\kappa^4} - \frac{b_3}{b_1}B_1 
+ \frac{119\tilde{V}_0^4B_0}{864000b_1\kappa^8}  
\nonumber \\
&-& \frac{\tilde{V}_0^2b_2B_0}{1800b_1\kappa^4} - \frac{19b_1\tilde{V}_0^2B_0}{54000\kappa^4} 
+ \frac{3\alpha B_0^3}{32b_1\kappa^2}\,.
\label{bb2}
\end{eqnarray}
%
%
The particular solution is
\begin{eqnarray}
\label{three}
R_{3p} &=& c_{31}\cos(\kappa\xi) + c_{32}\cos(3\kappa\xi) + c_{33}\sin(3\kappa\xi) 
\nonumber \\
&+& c_{34}\sin(\kappa\xi) 
+ c_{35}\cos(4\kappa\xi) + c_{36} 
\nonumber \\
&+& c_{37}\sin(4\kappa\xi) + c_{38}\cos(5\kappa\xi) 
+ c_{39}\cos(\kappa\xi) 
\nonumber \\
&+& c_{310}\sin(5\kappa\xi) + c_{311}\sin(\kappa\xi)\,,
\end{eqnarray}
where
\begin{eqnarray}
c_{31} = &-&\frac{7b_1^2\tilde{V}_0A_0}{54\kappa^2} + \frac{b_2\tilde{V}_0A_0}{9\kappa^2} 
- \frac{11\tilde{V}_0^3A_0}{4320\kappa^6} \nonumber \\
&+& \frac{b_1\tilde{V}_0A_1}{9\kappa^2} - \frac{\tilde{V}_0A_2}{6\kappa^2}\,,
\end{eqnarray}
\begin{eqnarray}
c_{32} &=& \frac{31b_1^2\tilde{V}_0A_0}{250\kappa^2} - \frac{3b_2\tilde{V}_0A_0}{25\kappa^2} 
+ \frac{17\tilde{V}_0^3A_0}{12000\kappa^6} \nonumber \\
&-& \frac{3b_1\tilde{V}_0A_1}{25\kappa^2} + \frac{\tilde{V}_0A_2}{10\kappa^2}\,,
\end{eqnarray}
\begin{eqnarray}
c_{33} &=& \frac{17\tilde{V}_0^3B_0}{12000\kappa^6} - \frac{3b_1\tilde{V}_0B_1}{25\kappa^2} + 
\frac{31b_1^2\tilde{V}_0B_0}{250\kappa^2} 
\nonumber \\
&-& \frac{3b_2\tilde{V}_0B_0}{25\kappa^2} + \frac{\tilde{V}_0B_2}{10\kappa^2}\,, 
\end{eqnarray}
\begin{eqnarray}
c_{34} &=& \frac{b_2\tilde{V}_0B_0}{9\kappa^2} - \frac{7b_1^2\tilde{V}_0B_0}{54\kappa^2} - 
\frac{\tilde{V}_0B_2}{6\kappa^2} 
\nonumber \\
&+&  \frac{b_1\tilde{V}_0B_1}{9\kappa^2} - \frac{11\tilde{V}_0^3B_0}{4320\kappa^6}\,,
\end{eqnarray}
\begin{align}
c_{35} &= -\frac{19b_1\tilde{V}_0^2A_0}{1800\kappa^4} + \frac{\tilde{V}_0^2A_1}{240\kappa^4}\,, \quad
  c_{36} = \frac{\tilde{V}_0^2A_1}{48\kappa^4} - \frac{b_1\tilde{V}_0^2A_0}{72\kappa^4}\,, \notag \\
  c_{37} &= -\frac{19b_1\tilde{V}_0^2B_0}{1800\kappa^4} + \frac{\tilde{V}_0^2B_1}{240\kappa^4}\,, \quad
  c_{38} = \frac{\tilde{V}_0^3A_0}{10080\kappa^6}\,, \notag \\
  c_{39} &= -\frac{\tilde{V}_0^3A_0}{288\kappa^6}\,, \quad
  c_{310} = \frac{\tilde{V}_0^3B_0}{10080\kappa^6}\,, \quad
  c_{311} = \frac{\tilde{V}_0^3B_0}{288\kappa^6}\,. 
\end{align}
%
%
Similar to what occurs at $O(\varepsilon^2)$, the coefficient $c_{36}$ is the prefactor for $\cos(0\xi)$ and a $\sin(0\xi)$ term (not shown) occurs in (\ref{three}) as well. The extra terms (from the resonance relation) that go into the slow evolution equations and the resulting expressions for the periodic orbits (i.e., the equilibria of the slow flow) arise from the terms with prefactors $c_{39}$ and $c_{311}$.  (The harmonics corresponding to the coefficients $c_{31}$ and $c_{34}$ are always secular, but those corresponding to $c_{39}$ and $c_{311}$ are secular only for 1:2 superharmonic resonances.)

One inserts equilibrium values of $A_j$ and $B_j$ ($j \in \{0,1,2,3\}$) into 
Eqs.~(\ref{zeroB}), (\ref{1B}), (\ref{2B}), and (\ref{3B}) to obtain the spatial
profile $R = R_0 + \varepsilon R_1 + \varepsilon^2 R_2 + \varepsilon^3 R^3 + O(\varepsilon^4)$ 
used as the initial wavefunction in numerical simulations of the GP equation (\ref{ph1}) with a stationary OL.


\begin{thebibliography}{99}
\bibitem{note1} The small sizes of the lattice depth and nonlinearity coefficient allow us to 
perturb from base states consisting of trigonometric functions (regular harmonics). 
One can relax these conditions using the much more technically demanding approaches of perturbing 
from either Mathieu functions (removing the restriction on the size of the amplitude of the 
periodic potential) or elliptic functions (removing the restriction on the size of the nonlinearity coefficient).
\bibitem{note2} We use the same values of $A_0$ and $B_0$ (the prefactors of the harmonics in $R_0$; 
see the appendices) as for the half-period states, even though these parameters are not 
constrained for quarter-period solutions. As discussed in Appendix A, they are constrained for half-period states.
\end{thebibliography}
\end{document}